\documentclass[aps,twocolumn,preprintnumbers,eqsecnum,bibnotes,showpacs]{revtex4}
\usepackage[tbtags]{amsmath}
\usepackage{graphicx}
\usepackage{bm}

\def\ts{\textstyle}
\def\half{\ts\frac{1}{2}}

\def\beq{\begin{equation}}
\def\eeq{\end{equation}}
\def\eeql#1{\label{#1} \end{equation}}

\def\C{\Gamma}
\def\D{\Delta}
\def\W{\Omega}
\def\a{\alpha}
\def\b{\beta}
\def\c{\gamma}

\def\e{\epsilon}
\def\w{\omega}
\def\hn{\mskip-0.5\thinmuskip}
\def\hp{\mskip0.5\thinmuskip}
\def\im{\mathop{\rm Im}\nolimits}
\def\re{\mathop{\rm Re}\nolimits}
\def\Ei{\mathop{\rm Ei}\nolimits}
\def\res{\mathop{\rm res}\nolimits}
\newcommand{\To}{\rightarrow}

\begin{document}

\title{Linearized Perturbations of a Black Hole: Continuum Spectrum}

\author{P.T. Leung}
\author{Alec \surname{Maassen van den Brink}}
\thanks{Corresponding author;\\ electronic address: \texttt{alec@dwavesys.com}}
\author{K.W. Mak}
\author{K.~Young}
\affiliation{Physics Department, The Chinese University of Hong Kong, Hong Kong, China}

\date{\today}

%==========================================================

\begin{abstract}
Linearized perturbations of a Schwarzschild 
black hole are described, for each angular momentum~$\ell$,
by the well-studied discrete  
quasinormal modes (QNMs), and in addition a continuum. 
The latter is characterized by a cut strength $q(\c{>}0)$ 
for frequencies $\w = -i\c$.  We show that  
(a)~$q(\c{\downarrow}0) \propto \c$, (b) $q(\C)=0$
at $\C=(\ell{+}2)!/[6(\ell{-}2)!]$,
and (c) $q(\c)$ oscillates with period $\sim 1$ ($2M\equiv1$).
For $\ell=2$, a pair of QNMs are found beyond the cut on the
unphysical sheet very close to $\C$, leading to
a large dipole in the Green's function \emph{near} $\C$.
For a source near the horizon and a distant observer,
the continuum contribution relative to that of the QNMs is small.
\end{abstract}


\maketitle

%==========================================================

\section{Introduction}
\label{sect:intro}

Gravitational waves propagating on a Schwarzschild background 
probe the nontrivial spacetime around the event horizon. 
If and when detected~\cite{LIGO}, their signature may 
confirm that black holes exist. 
Because waves escape to infinity
and into the horizon, the system is dissipative,
and described by its spectrum in the lower half frequency plane---of interest
both for signal interpretation and in its own right.
The discrete quasinormal modes (QNMs) have been 
thoroughly studied \cite{Chandra,Ferrari,Nollert,Liu}; 
this paper characterizes the continuum, 
about which little is hitherto known.

For a black hole of mass $M$ (below 
$c = G = 2M=\nobreak1$) and each angular momentum $\ell$,
the radial functions $\psi$ of scalar ($s{=}0$) or electromagnetic ($s{=}1$) waves 
and axial gravitational perturbations ($s{=}2$) are governed by 
a generalized Klein--Gordon or so-called Regge--Wheeler equation (KGE or RWE) 
$[d_x^2 + \w^2 - V(x)] \psi(x,\w)=0$; $x=r+\ln (r{-}1)$ is the tortoise coordinate and $r$ the 
circumferential radius. The potential 
\beq
  V(r)=\left( 1 - \frac{1}{r} \right) \left[ \frac{\ell(\ell{+}1)}{r^{2}} +
  \frac{1{-}s^2}{r^3} \right],
\eeq
describing scattering by the background~\cite{R-W}, behaves as 
$V_{\ell}(x{\To}\infty)
\equiv V(x)-\ell(\ell{+}1)/x^{2} \sim 2\ell(\ell{+}1)\ln x/x^{3}$ 
and $V(x{\To}{-}\infty) \sim \left[ \ell(\ell{+}1)+1-s^{2} \right] e^{x-1}$.
We impose outgoing-wave conditions (OWCs)
$\psi(x{\To}{\pm}\infty,\w) \sim e^{i\w |x|}$. 
(For $x\to-\infty$, waves thus go into the horizon.)

Polar gravitational perturbations are governed by 
the Zerilli equation (ZE)~\cite{Zerilli}, which is the KGE with the 
potential $\tilde{V}(x)=V(x)+2d_xW(x)$, where
\beq
  W(r) = \C + \frac{3(r-1)}{r^{2}(2\nu r+3)}\;,
\eeql{eq:suppot}
$\C=(\ell{+}2)!/[6(\ell{-}2)!] = \frac{2}{3}\nu(\nu{+}1)$, and $\nu=\frac{1}{2}(\ell{-}1)(\ell{+}2)$.
In fact, $W(x) = -g'(x,i\C) / g(x,i\C)$ [cf.\ below (\ref{J12})];
$g(x,\omega)$ is defined in general below.
The solutions $\psi$ of the RWE and 
$\tilde{\psi}$ of the ZE are 
related by ``intertwining'' or supersymmetry (SUSY): $\tilde{\psi}(x,\w) = [ d_x + W(x) ] \psi(x,\w)$ \cite{Chandra,chand,Wong}. Thus, also the two continua are closely related.

\begin{figure}
  \includegraphics[height=3in,angle=270]{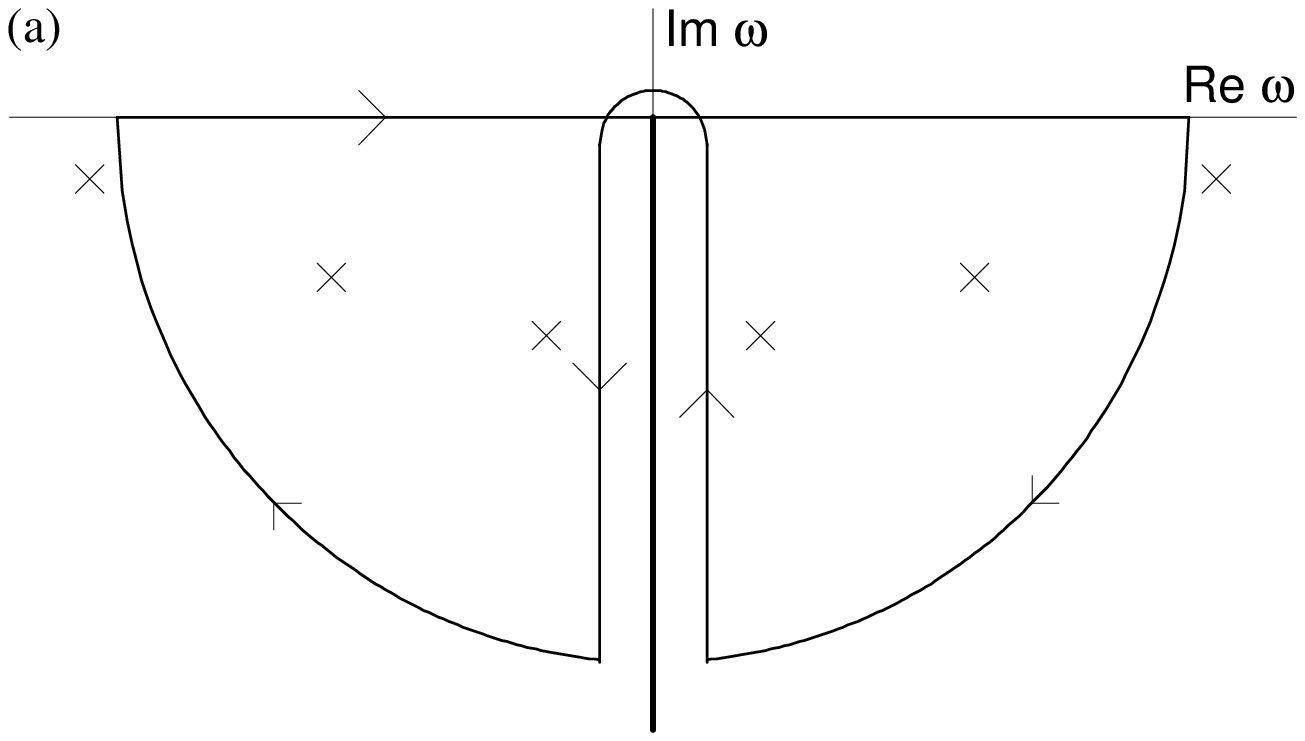}

  \vspace{1mm}
  \includegraphics[height=3in,angle=270]{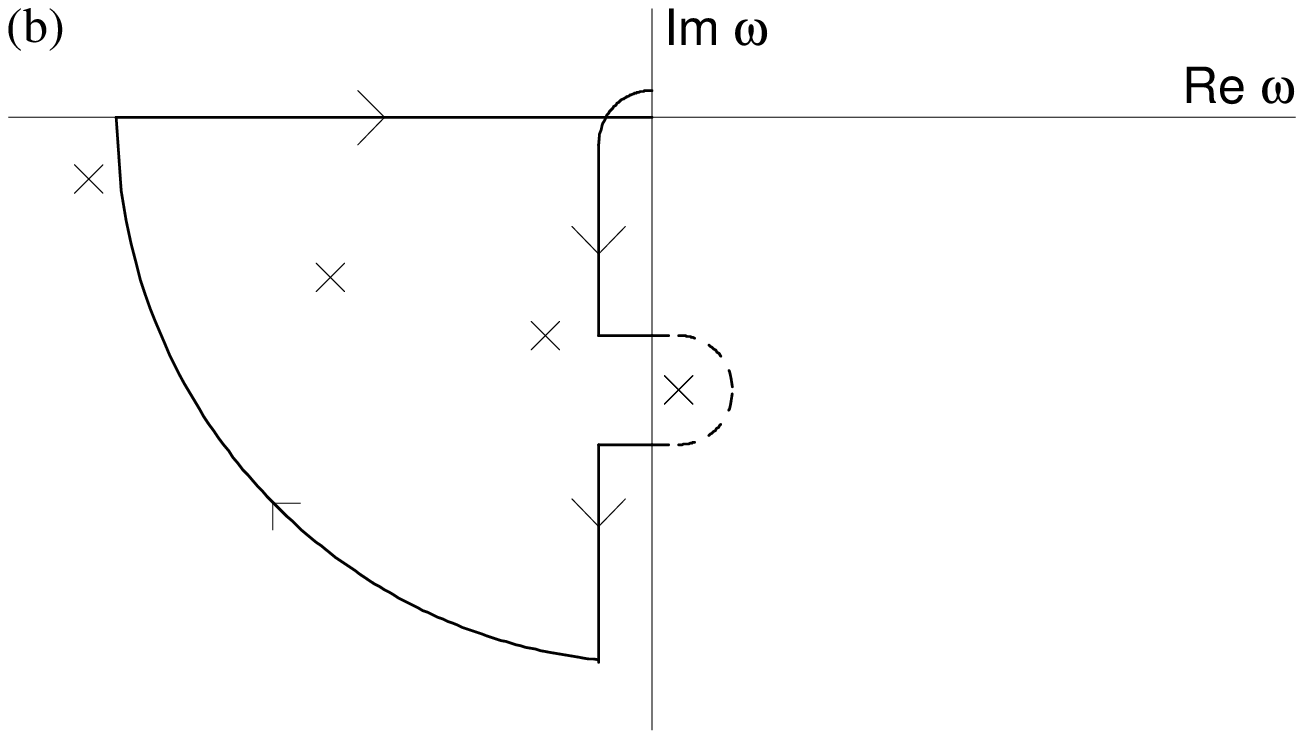}
  \caption{Fourier inversion of $\bar{G}$; crosses denote poles in the $\w$\protect\nobreakdash-plane.
(a)  Conventional contour involving a cut
on the NIA.
(b)  Modified contour detouring around the unconventional pole near $\C$ 
(contour on unphysical sheet shown as broken line).
For simplicity the mirror contour is not shown.}
  \label{fig1}
\end{figure}

The signal $\psi(x,t {\ge} 0)$ depends on $\{\psi(y,0),\dot{\psi}(y,0)\}$ 
through the Green's function 
$G(x,y;t) = 
\int (d\w /2\pi)\*\bar{G}(x,y;\w)e^{-i\w t}$. 
Closing the contour in the lower half $\w$\nobreakdash-plane 
(Fig.~\ref{fig1}a) separates $\bar{G}$ into (a)~the large semicircle ($|\w| \To \infty$), giving a prompt signal propagating directly from $y$ to $x$ and vanishing after a finite~$t$ \cite{Ching-cut,Bachelot};
(b)~the QNM poles, giving rise to a ringing signal dominating at intermediate~$t$ \cite{Leaver,Ching-QNM}; and
(c)~our main focus, the cut on the negative imaginary axis (NIA) $\w=-i\c$:
\beq
  \D \bar{G}(x,y;-i\c) =\bar{G}_{+}(x,y;-i\c)-\bar{G}_{-}(x,y;-i\c)\; ,
\eeql{eq:cut}
where $\bar{G}_{\pm} (-i\c)=\lim_{\e \downarrow 0} \bar{G}(-i\c {\pm}\e)$ are continuations from $\pm\w>0$. The physical sheet for $\bar{G}_{+}$ ($\bar{G}_{-}$) lies to the right (left) of the NIA.  [However, the opposite sides are unphysical only for the conventional choice of the
cut, as in (\ref{eq:cut}) and Fig.~\ref{fig1}a.]  
The continuum is given by $\D \bar{G}$, and for $\c \To 0$ causes the late-$t$ behavior \cite{Leaver,Ching-cut}.

In general, the Green's function $\bar{G}(x,y)=\bar{G}(y,x)$ is
\beq
  \bar{G}(x,y;\w)=\frac{f(y,\w)g(x,\w)}{J(g,f;\w)} \; , \qquad y<x \; ,
\eeql{eq: green}
where $f$ ($g$) solves the KGE with the left (right) OWC, and $J(g,f;\w)=gf'-fg'$ is their Wronskian. Although $\bar{G}$ is normalization-independent, for definiteness we define $f(x{\rightarrow}{-}\infty,\w) \sim 1 \cdot e^{-i\w x}$ and $g(x{\rightarrow}\infty, \w) \sim 1 \cdot e^{i\w x}$ (adopted for $\im\w\ge0$, and continued to $\im\w<0$).
%We consider a source near the horizon (say $y < 0$) and a distant observer (say $x > 0$).
At a zero of~$J$, $f \propto g$ satisfies both OWCs and defines a QNM. These are well understood, and we turn to cuts.

If $V$ has its support in say $[-d,d]$, 
the OWCs can be imposed at $\pm d$, 
so the KGE is integrated over a finite distance; hence, $f,g$ are analytic in $\w$. 
The same holds if $V$ 
decays faster than exponentially. 
If however $V(x{\To}{-}\infty) \sim \sum_{k}v_{k}e^{\lambda kx}$
(here $\lambda=\nobreak1$), typically $f$ has poles 
(``anomalous points") at $\w_n=-in\lambda /2$ (readily shown by Born 
approximation, i.e., a power series in~$e^x$). 
These are 
removable by scaling $f(\w) \mapsto \chi(\w) = 
(\w{-}\w_n)f(\w)$, leaving $\bar{G}$ unaffected~\cite{anom}.
However, $\{v_k\}$ could conspire to ``miraculously" make some $\w_n$ non-singular.  
For the RWE, $n = N \equiv 2\C$ ($= 8$ for $\ell =2$) is miraculous.
Although miracles can be studied by finite-order Born approximation, here an exact solution exists at $\C$ [cf.\ (\ref{eq:suppot}) and~(\ref{def-xi1})], 
the only miraculous point (for any~$\ell$)~\cite{Alec}.

On the other hand, for $x \To +\infty$, the centrifugal barrier does not scatter and has no real effects~\cite{Ching-cut}, so we should consider
the next asymptotic term
\beq
  V_{\ell}(x {\To} \infty) \sim \frac{\ln x}{x^3}
  = \int_{0}^{\infty}\!\!\! d\lambda\, (3{-}2\c_\mathrm{E}{-}\ln\lambda)
  \lambda^{2}e^{-\lambda x}
\eeql{Vtail}
($\c_\mathrm{E}$ is Euler's constant). The superposition of exponentials spreads the poles at $-in\lambda /2$ into a cut due to the power-law tail.
In $\bar{G}$ in (\ref{eq: green}), only $g$ is discontinuous and we 
are led to study $\D g(-i\c) \equiv g_+(-i\c)-g_-(-i\c)$, cf.\ (\ref{eq:cut}).
Section~\ref{sect:cut} analyzes this in terms of a 
position-independent cut strength~$q(\c)$. The result is  
checked against
(a)~the $\c \To 0$ limit, and 
(b) the zero and known slope at $\c = \C$~\cite{Alec}. 

Section~\ref{sect:Green} deals with the Green's function, 
in particular $\D \bar{G}$ and 
the limiting function $\D \bar{G}^{\rm L}$ describing
propagation from near the horizon to infinity.
Of course $\D \bar{G} \propto q(\c)$ has a zero
at $\c = \C$; surprisingly, there is a large
contribution, approximately a dipole, {\em near\/} $\C$, attributable to a pair of nearby QNMs on the unphysical sheet~\cite{bhL}. These poles are absent in $q$ itself, and point to a relation to $f$---in particular to $J$ in (\ref{eq: green}), which must have a zero at these positions. This leads to an analytic treatment, by linearization about $\Gamma$.
Furthermore the $\c \To 0$ 
behavior is examined to recover $G(x,y;t {\To} \infty)$.

Some elements of the analysis are computationally hard (CH),
in a precise sense: consider the evaluation of $g(x,-i\c)$ 
given that $g(z{\To}\infty,-i\c) \sim e^{\c z}$.
If the OWC is imposed at $x=L$, one needs accuracy $e^{-2\c L}$ to 
exclude an $O(1)$ admixture of the wrong solution. 
If $V$ does not have finite support ($L \To \infty$) no 
finite accuracy suffices, defeating direct integration of the KGE~\cite{Tam}. 
Instead, one must continue from the upper 
half $\w$-plane (where growing and decaying 
solutions are interchanged) to the lower one---implicit in all analytic formulas, e.g., the Born approximation or Leaver's series (see below). 
In contrast, evaluating a decaying function given its asymptotics, 
e.g.\ $g(z,i\c) \sim e^{-\c z}$, is 
not CH. Calculating $g(x,\w)$ from $\{g(z,\w),g'(z,\w)\}$ is also not 
CH if $|x-z|$ is finite. In Sections \ref{sect:cut} and~\ref{sect:Green}, handling the CH parts is the technical issue. For $z\rightarrow -\infty$, the potential tail is exponential,
and the calculation is {\it not\/} CH when matching to a finite-order Born approximation
instead of to $f(z, -i\c) \sim e^{-\c z}$~\cite{Tam}.

%===========================================================

\section{Cut strength}
\label{sect:cut}

%----------------------------------------------------------
\subsection{Definition and properties}
\label{subsect:cutdef}

On the NIA, $g_{\pm}\sim1 \cdot e^{\c x}$ satisfy 
the same RWE. Hence 
$\D g \sim 0 \cdot e^{\c x}$ is the small solution $\propto\nobreak g(+i\c)$. 
Since $g(-\w^*)=g^*(\w)$, $\D g$ is imaginary, 
so we introduce the real cut strength $q$~\cite{Leaver,Alec}, to be studied numerically:
\beq
  \D g(x,-i\c)=iq(\c)g(x,+i\c) \;.
\eeql{eq:defq}
Since $g$ is defined by the OWC at $x \rightarrow \infty$, (\ref{eq:defq}) defines $q$ independent of $V(x)$ at any finite $x$: if, say, $V_1(x{>}L) = V_2(x{>}L)$, the corresponding $q_1$ and $q_2$ are identical. Leaver~\cite{Leaver} has given a formal expression for~$q$, which is however nontrivial to evaluate.

As a result of the SUSY relationship between
the RWE and ZE, the latter's cut strength $\tilde{q}$ obeys~\cite{Alec}
\beq
  (\C+\c)\hp\tilde{q}(\c) = (\C-\c)\hp q(\c) \;.
\eeql{eq: q-q}

%----------------------------------------------------------
\subsection{Numerical evaluation}
\label{cut-num}

\begin{figure}
  \includegraphics[height=3in,angle=270]{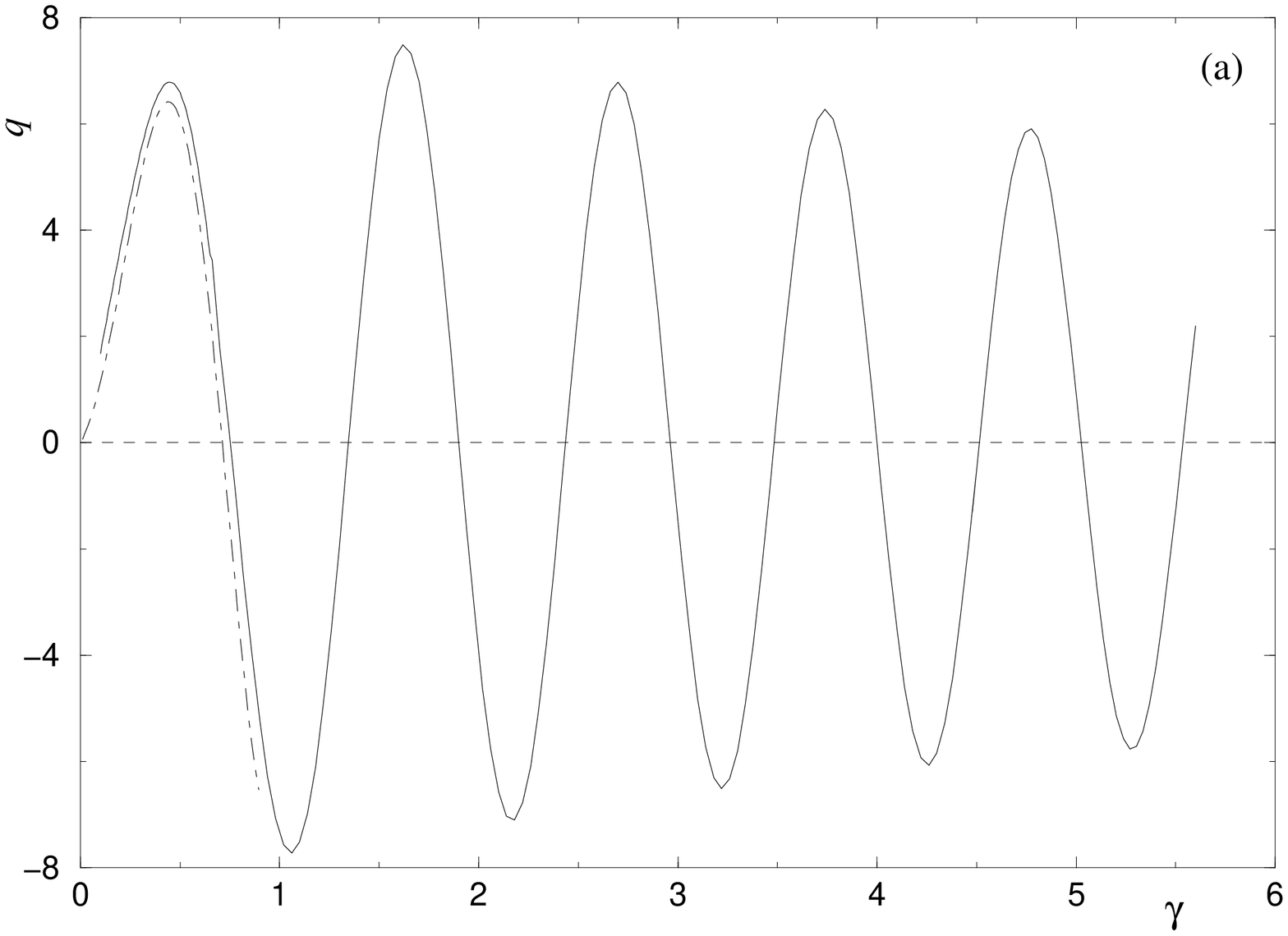}

  \vspace{1mm}
  \includegraphics[height=3in,angle=270]{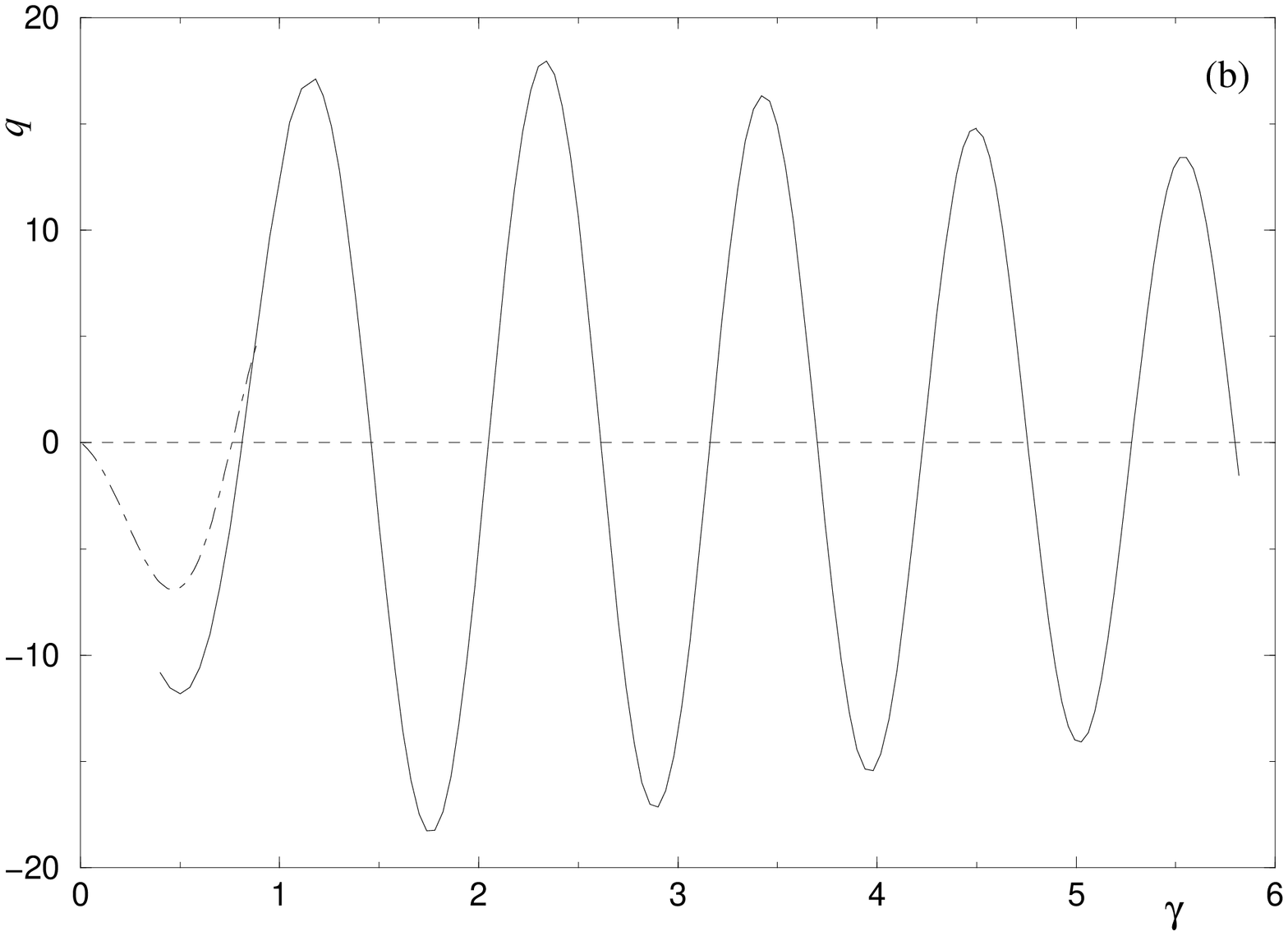}

  \vspace{1mm}
  \includegraphics[height=3in,angle=270]{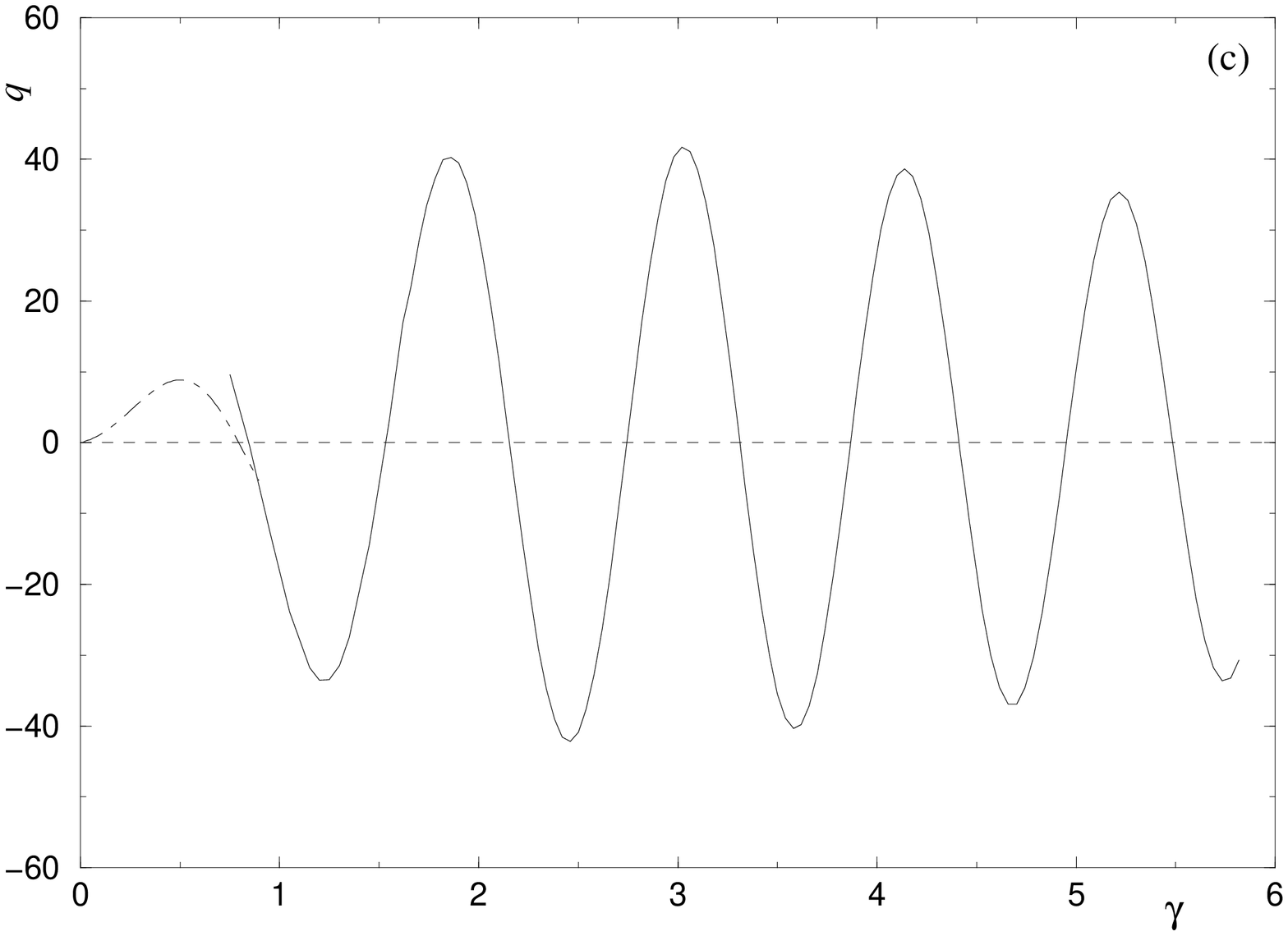}
  \caption{Plots of $q(\c)$ for (a) $\ell=2$, (b) $\ell=3$, and (c) $\ell=4$.
The solid (broken) lines are the numerical result (analytic approximation).}
  \label{fig2}
\end{figure}

Since $g(x,+i\c)$ is not CH, it is simply integrated from large $x$. We compute $g(x,-i\c {\pm} \e)$ by Miller's algorithm (failing at $\e=0$), taking $\e \downarrow 0$ in the difference for $\D g$; see Appendix~\ref{app:NA}~\cite{clarify}. The solid lines in Fig.~\ref{fig2} show the ensuing $q(\c)$ for $\ell=2,3,4$, and are key results. 

For $\ell=2$, $q(\c)=0$ at $\c=$ 0.75, 1.35, 
1.90, 2.44, 2.96, 
3.48, 4.00, 4.51, 5.03, 5.54, \ldots, suggesting that the spacing
approaches $\frac{1}{2}$. Indeed, by a WKB analysis~\cite{Alec2},
\beq
  q(\c) \sim 4 \cos(2\pi\c) + O(\c^{-1/2})\; ,
\eeql{eq:asmp}
where only the correction depends on~$\ell$. The zero at $\C$ and moreover $q'(\C) \approx -37.6$ agree with~\cite{Alec}
\begin{align}
  q'(\C)&=-\frac{45\pi}{137438953472} \left( 2100027e^{8}+30148389005 \right)\notag\\
  &= -37.45\ldots \;,\label{eq:q-slope}
\end{align}
confirming both (\ref{eq:q-slope}) and our numerical accuracy.  

For $\ell = 3,4$, $q(\gamma)$ is similarly
oscillatory, but the range of validity does not include $\C$ ($=20$ and 60 respectively).

For $\ell\To\infty$, one has $q'(\C) \sim -\sqrt{96\pi^3}$~\cite{q'}, of the same order as
(\ref{eq:q-slope}). Thus, in each of the analytically accessible
regimes $\c\downarrow0$ [below (\ref{eq: approx of q})], $\c\To\C$, and
$\c\To\infty$ [cf.\ (\ref{eq:asmp})], $q(\c)$ has a
finite limiting behavior for large~$\ell$, consistent with no new
features developing in these curves.

%----------------------------------------------------------
\subsection{Analytic approximation}
\label{cut-approx}

Small $\c$ relates to late-$t$ behavior, controlled by (a)~many finite-$r$ scatterings, whose effect vanishes exponentially, and (b)~a few large-$r$ scatterings, which therefore dominate. In terms of $u(r) \equiv \sqrt{1-1/r}\hp\psi(r)$, the RWE reads $d_r^2u + P u = 0$~\cite{Murphy}, where $P(r) = [r/(r{-}1)]^2[\w^2+r^{-3}-\frac{3}{4}r^{-4}-V(r)]$; the ZE follows in complete analogy. For large~$r$, both $P$ and $\tilde{P}$ can be approximated by $P_0 (r) = \w^{2} + 2\w^2r^{-1}+[3\w^2-\ell(\ell{+}1)]r^{-2}$, with errors $P(r) -P_0(r) \sim b_3 r^{-3}$, $\tilde{P}(r) -P_0(r) \sim \tilde{b}_3 r^{-3}$. For $s{=}2$, $b_3 = 4\omega^2 - 2\nu +2<\nobreak0$
for small $\omega$, so $\tilde{b}_3 = b_3 + 6/\nu$ has the smaller magnitude. Thus we regard the equation with $P_0$ as an approximation for the ZE, and obtain results for the RWE through SUSY~(\ref{eq: q-q}).

Re-expressing the above in terms of $\psi$, the leading $\tilde{V}_\ell(x)\sim x^{-3} \ln x$ is reproduced exactly. The error is~$\sim\nobreak x^{-3}$, apparently down by only a factor $\ln x \sim \ln t$, but a pure $x^{-3}$ term does \emph{not} generate a late-$t$ tail to first Born approximation~\cite{Ching-cut}. Thus the leading correction is an extra power of $x^{-1} \sim t^{-1}$ (up to logarithms).

The equation with $P_0$ is the hydrogen problem with fractional angular momentum. The OWC for $r\To\nobreak\infty$ selects its confluent-hypergeometric solution~\cite{Slater} and the branch of $\sqrt{-\omega^2}$, giving $\tilde{u}(r,\w)\approx(-2i\w)^{\sigma{-}i\w}\*r^\sigma\* e^{i\w r}\*U(\sigma{-}i\w,2\sigma;-2i\w r)$ with $2\sigma \equiv 1+\sqrt{(2\ell{+}1)^2-12\w^2}$.

Since we consider $\c \To 0$ at fixed $r$,  
$-i\w r \approx 0$, so
\beq
  \tilde{q}(\c) = \frac{\D \tilde{u}(r,-i\c)}{i\tilde{u}(r,i\c)}
  \approx -2\frac{\C(\sigma{+}\c)}{\C(\sigma{-}\c)} \frac{\sin[\pi(\sigma{+}\c)]}{(2\c)^{2\c}}\;,
\eeql{eq: approx of q}
with $\C$ the Gamma function (not the special frequency). This approximation is shown by broken lines in Fig.~\ref{fig2}. In particular, it shows that $q'(0) = \tilde{q}'(0) = (-1)^{\ell} 2\pi $~\cite{Ching-cut}.

%===========================================================
\section{Green's function}
\label{sect:Green}

%---------------------------------------------------------
\subsection{Evaluation and general properties}

\begin{figure}
  \includegraphics[height=3in,angle=270]{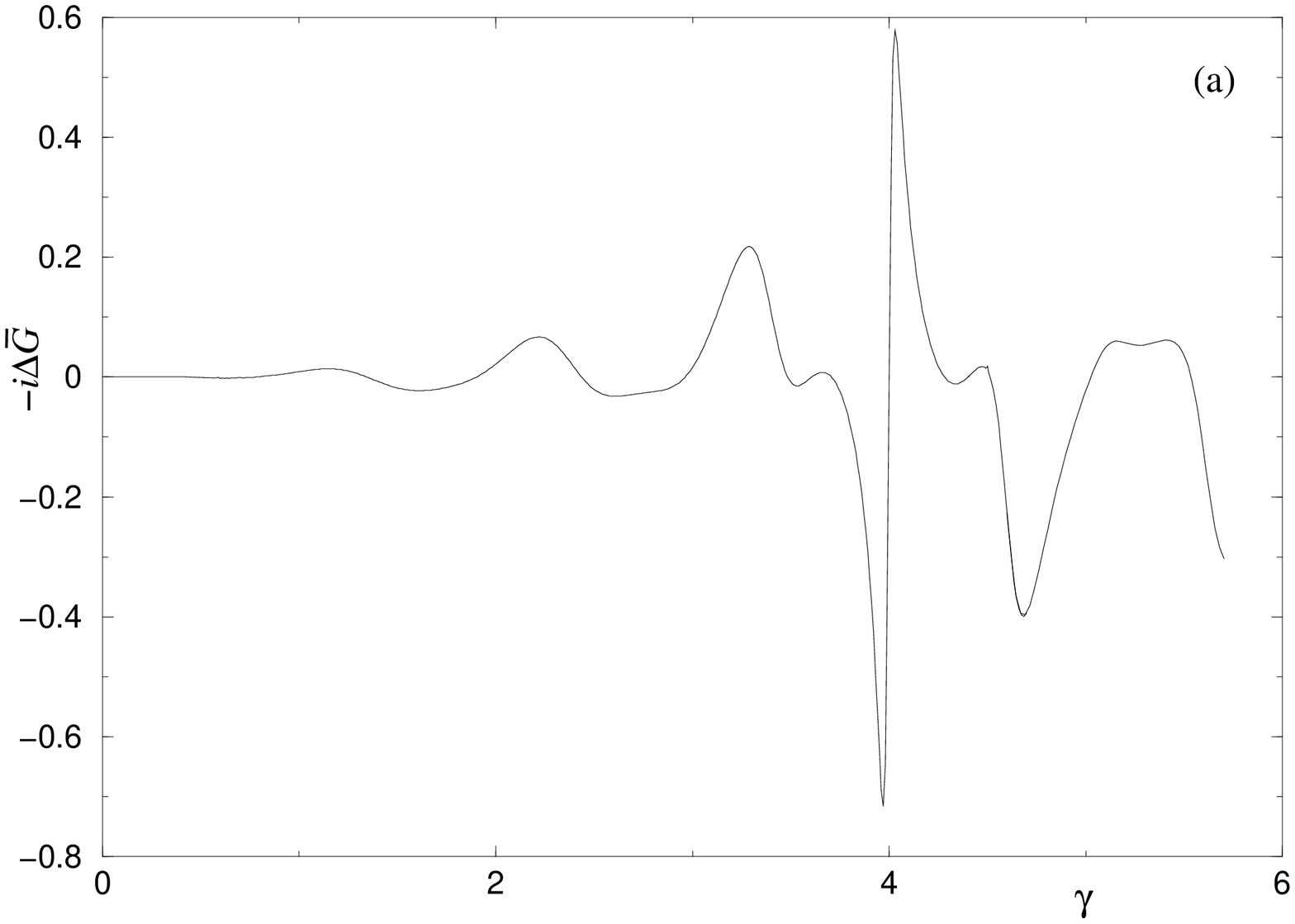}

  \vspace{1mm}
  \includegraphics[height=3in,angle=270]{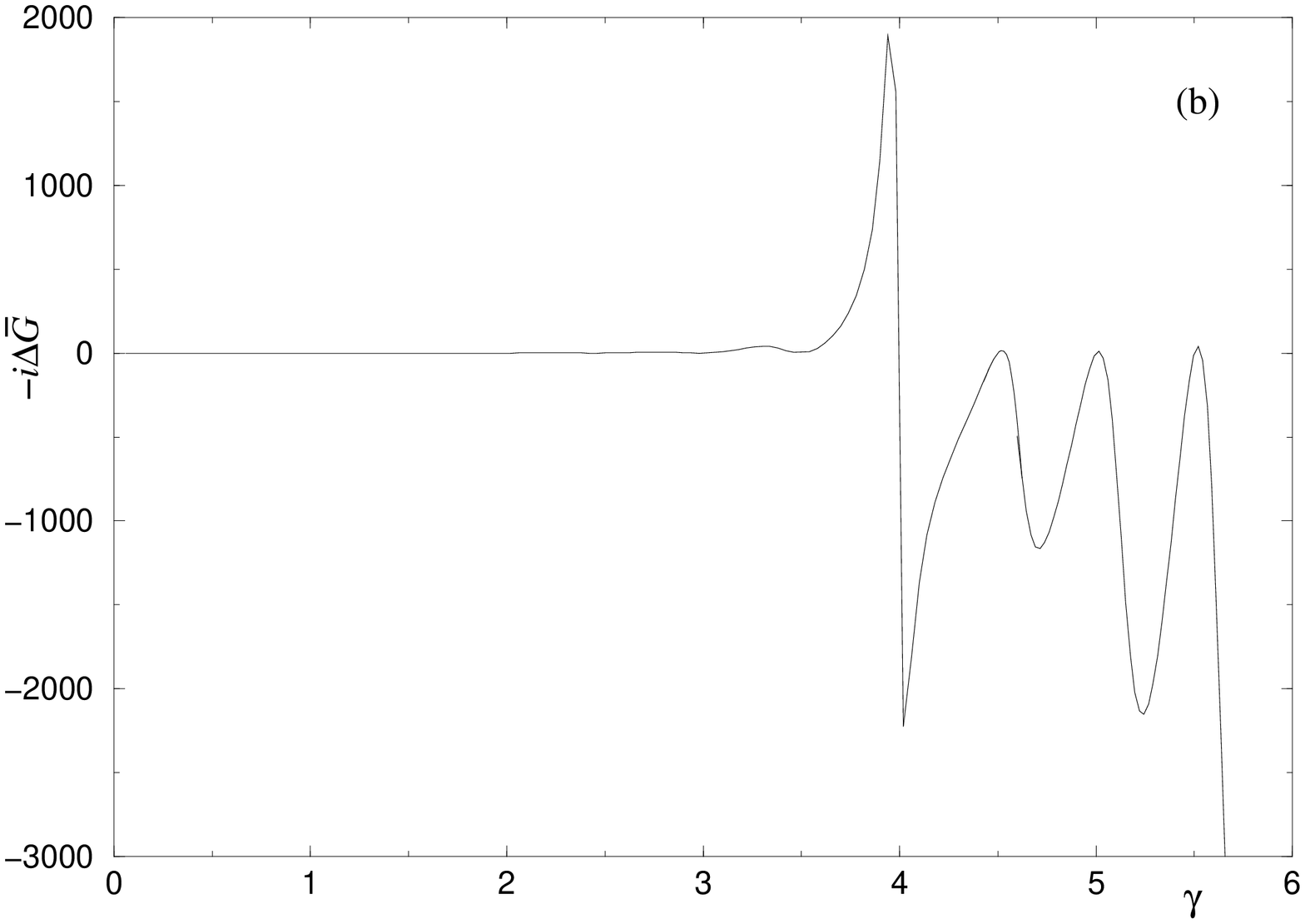}
  \caption{Plots of $-i \D \bar{G}(x,y;-i\c)$ for $\ell=2$ and
(a) $x=0.2$, $y=0.1$, (b) $x=1.0$, $y=-1.0$.}
  \label{fig4}
\end{figure}

Using $J(g_{-}, g_{+}; -i\c) = -2i\c q(\c)$, one finds
\beq
  \D \bar{G}(x,y;-i\c)
  = -2i\c q(\c) \frac{f(x,-i\c)f(y,-i\c)}{J_{+}(-i\c)  J_{-}(-i\c) } \; ,
\eeql{eq: DG and q-2}
with $J_\pm \equiv J(g_\pm, f)$. We evaluate $g_\pm$ as in Section~\ref{cut-num} and use Jaff\'e's series for~$f$ (Appendix~\ref{app:jaffe}).
Fig.~\ref{fig4} shows $-i \D \bar{G}(x,y;-i\c)$ for $\ell =\nobreak 2$% and different $x,y$
; the two contrasting cases reveal a strong $x$- and $y$\nobreakdash-dependence.
% The prefactor in (\ref{eq: DG and q-2}) leads to many zeros, including at $\c=0$ and $\C$.

The above is the clearest, but more efficient is to rewrite (\ref{eq: green}) as (suppressing $\w=-i\c {\pm} \e$, $\e \ne 0$):
\beq
  \bar{G}(x,y) = 
 \frac{[f(y)/f(z)] \, [g(x)/g(z)]}{ f'(z)/f(z) - g'(z)/g(z) }\; .
\eeq
In principle any $z$ will do. 
However, if $\pm z \gg 1$, both $L(z) \equiv f'(z)/f(z)$ 
and $R(z) \equiv g'(z)/g(z)$ approach $\pm i\w$, so the exponentially small denominator is hard to evaluate. Thus choose $z \sim O(1)$ and
(a)~compute $R(z)$ (Appendix~\ref{app:NA}).
This CH step is carried out only at \emph{one}~$z$. 
(b)~Calculate $f$ as above for $L(z)$.
(c)~Up to irrelevant normalization, 
$f(y)$ and $g(x)$ follow by integrating the KGE over finite distances, yielding~$\bar{G}$. (d)~Let $\e \downarrow 0$. We have verified that the result agrees with~(\ref{eq: DG and q-2}).

As said below (\ref{eq:cut}), $\D\bar{G}(\c{\approx}0)$ governs $G(t{\To}\infty)$.
Since $\c\downarrow0$ relates to $x \To \infty$~\cite{Ching-cut}, one can neglect $V_\ell$ for all factors in (\ref{eq: DG and q-2}) except 
$q(\c) \approx (-1)^{\ell}\* 2\pi \c$ (Section~\ref{cut-approx}), giving 
$f(x,-i\c) \propto x^{\ell+1}$ and 
$g_\pm(x,-i\c ) \approx [(2\ell)!/\ell!]\*(-2\c x)^{-\ell}e^{\c x}$. This yields $\D \bar{G} \approx 4\pi i[(2\ell{+}1)!!]^{-2}\*(-xy)^{\ell{+}1}\c^{2\ell+2}$, elegantly reproducing~\cite{Leaver,Ching-cut}
\beq
  G(x,y;t) \approx  2 \frac{(2\ell{+}2)!}{[(2\ell{+}1)!!]^{2}}
  \frac{(-xy)^{\ell+1}}{t^{2\ell+3}} \;.
\eeq
A potential tail 
$V_{\ell}(x) \sim x^{-\alpha} \ln x$
generically leads to $G \sim t^{-(2\ell+\alpha)} \ln t$,
but in some special cases (including the RWE),
the $\ln x$ factor does not lead to a $\ln t$ factor~\cite{Ching-cut}.

%----------------------------------------------------------
\begin{figure}
  \includegraphics[height=3in,angle=270]{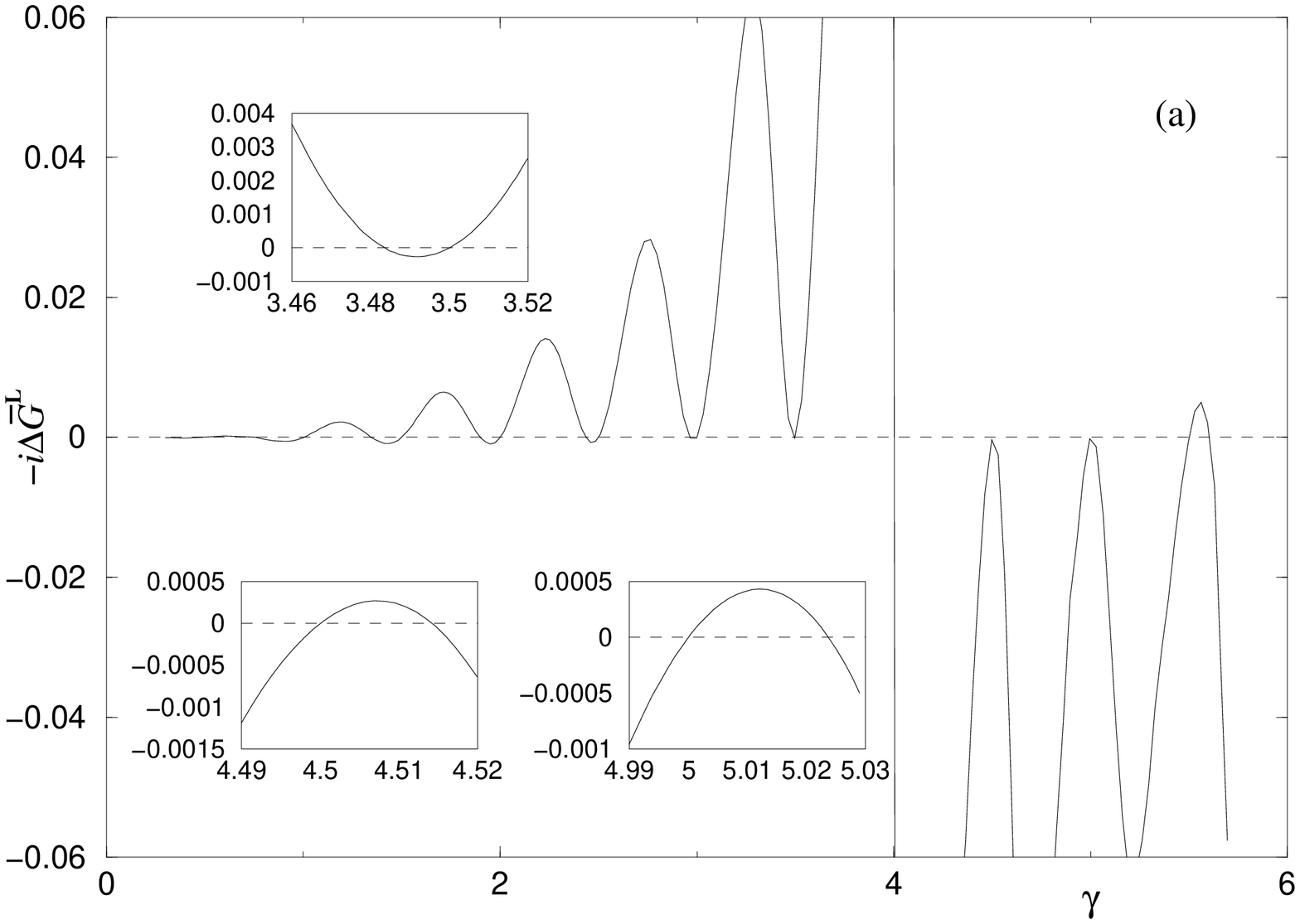}

  \vspace{1mm}
  \includegraphics[height=3in,angle=270]{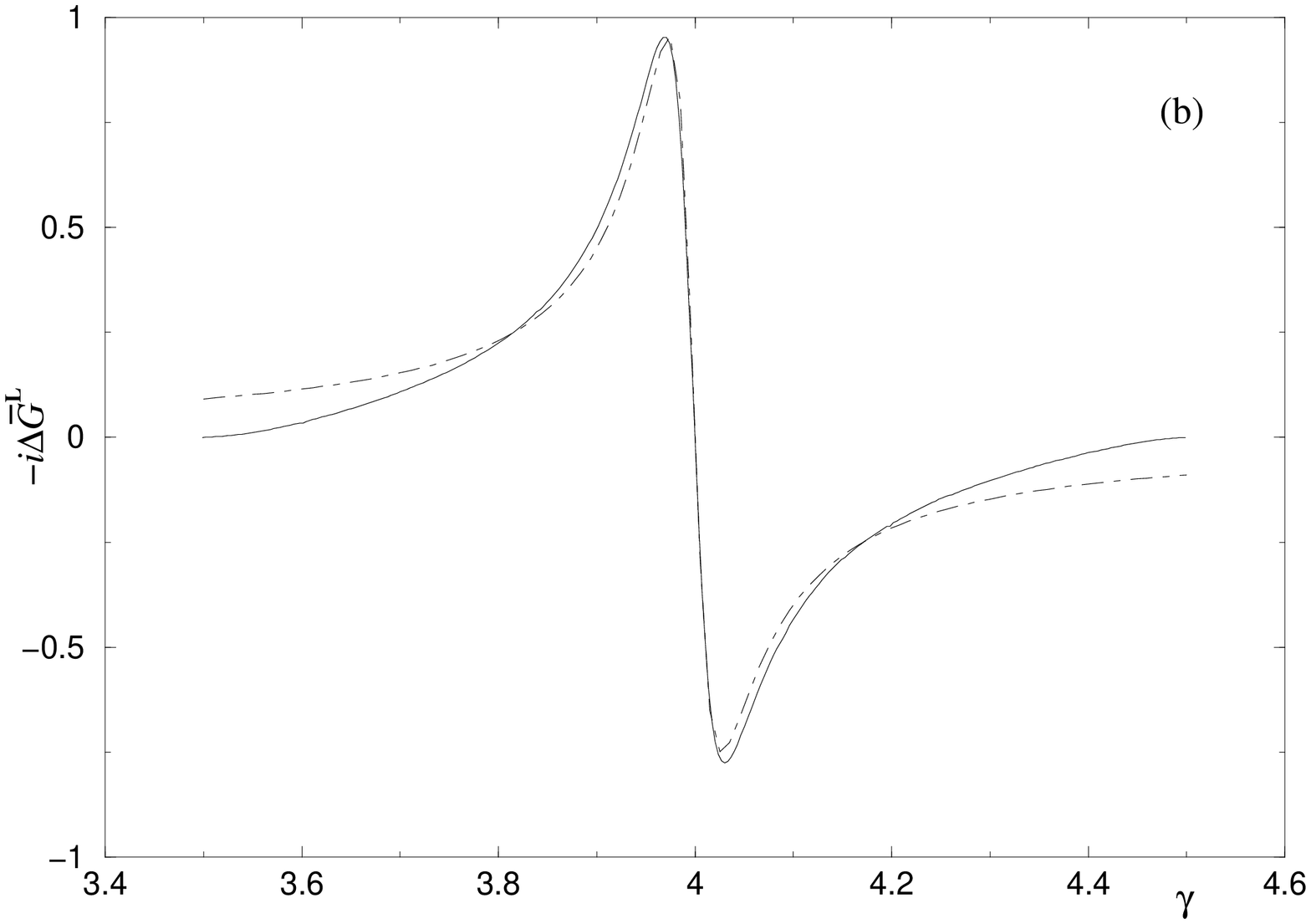}
  \caption{Plots of $-i \D \bar{G}^{\rm L}(-i\c)$ for $\ell=2$.
(a)  For $0 \le \c \le 5.6$.  Insets show some regions
with two close zeros.
(b)  The region $\c \approx 4$ expanded.
The solid line is the numerical result, and the broken line
is a fit to (\ref{eq:fit1}) with $a_2=-0.0227$.}
  \label{fig5}
\end{figure}

\subsection{Limiting $x$ and $y$}
\label{subsect:lim}

The physically important limit is $-y,x \rightarrow \infty$.  
Because $f(y, -i\c) \sim e^{-\c y}$ and $f$ in general also has
an outgoing part to the right [$f(x, -i\c) \sim e^{\c x}$],
there is a steep position dependence.
This is simply a result of the long signal
propagation time, and is removed if $t$ is measured 
from the first arrival at $t_0(x,y) \equiv x-y$.
[Closing the contour in the upper 
$\w$-plane readily shows that $G(t{<}t_0)=0$.]  Thus, consider
\beq
  G^{\rm L}(t') \equiv \lim_{x, -y \rightarrow \infty} G(x,y; t_0 {+} t') \; ,
\eeql{eq:cutint01}
with Fourier transform $\bar{G}^{\rm L}(\w)=J(g,f;\w)^{-1}$
by (\ref{eq: green}) and 
the normalization of $f,g$.
The cut is
\begin{gather}
  \D G^{\rm L} (t') = \int_0^{\infty} \frac{d\c}{2\pi i} \hp
  \D \bar{G}^{\rm L} (-i\c)\hp e^{-\c t'}\;,\label{eq:cutint03}\displaybreak[0]\\
  \D \bar{G}^{\rm L} (-i\c) = \Delta\! \left[ J(g,f;-i\c)^{-1} \right]\label{eq:cutint04a}
\end{gather}
[cf.\ (\ref{eq:cut}) for $\Delta$]. Results are shown in Fig.~\ref{fig5}.

\begin{figure}
  \includegraphics[height=3in,angle=270]{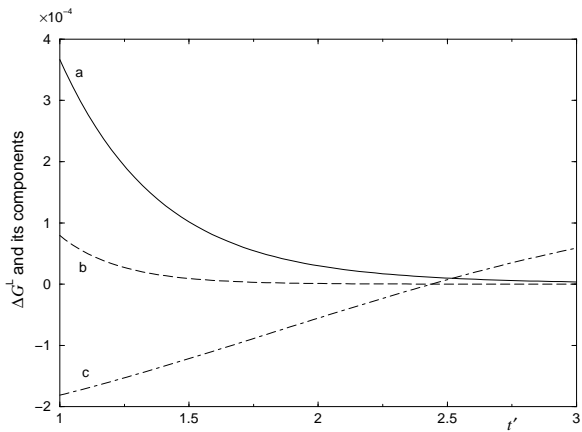}
  \caption{Solid line (a): continuum contribution $\D 
G^{\rm L}(t')$ for $\ell=2$. Dashed line (b): contribution of the pair of nearby unconventional poles. Dot-dashed line (c): corresponding contribution from the first QNM, scaled down by $10^3$.}
  \label{fig6}
\end{figure}

Although $\D \bar{G}^{\rm L}(-i\C)=0$,
surprisingly $\D \bar{G}^{\rm L}$ is largest \emph{near}
$\c = \C$, where it is approximately a dipole (Fig.~\ref{fig5}b; the dominance is
less pronounced if the source is displaced from
the horizon, cf.\ Fig.~\ref{fig4}).  The solid line in Fig.~\ref{fig6} shows $\D G^{\rm L}(t')$
[obtained by integrating (\ref{eq:cutint03}) up to
$\c = 5.60$, hence accurate except for very small~$t'$].

The cut $\D \bar{G}^{\rm L} (-i\c)$ vanishes at (a) the zeros of $q(\c)$ [cf. (\ref{eq: DG and q-2})], and (b)~$\c = \frac{1}{2}, 1, \frac{3}{2}, \ldots$.  The former depend \emph{only} on $V(x{\rightarrow} {+}\infty)$ [cf.~below (\ref{eq:defq})], the latter \emph{only} on $V(x{\rightarrow} {-}\infty)$, scaling with $\lambda$. If the two tails are separately adjusted the sequences are independent, but for the present $\lambda=1$, sequence~(a) acquires one integer member at $\c=\C$. However, this member simultaneously \emph{disappears} from sequence~(b) [cf.\ above (\ref{Vtail})], consistent with the \emph{first}-order zero at $\c=\C$ in Fig.~\ref{fig5}.

Some members of the two sequences are close, cf.\ the insets in Fig.~\ref{fig5}a. The many zeros also ensure that $\D \bar{G}^{\rm L}$ is generically small. (Section~\ref{subsect:compare} contains quantitative estimates.)

%----------------------------------------------------------
\subsection{Poles on unphysical sheet: numerics}

The behavior of $\D \bar{G}^{\rm L} (-i\c)$ for 
$\c \approx \C$ can in fact be
attributed to a pair of nearby QNM poles~$\w_\pm$.  Suppose $\bar{G}^{\rm L}_{\pm}(\omega)\approx (a_1 {\pm}ia_2)/(i\omega{-}4{-}b{\pm}ic)$, where $a_1, a_2, b$ are real and $c>0$
so $\w_\pm$ are on the unphysical sheets. In terms of $\xi=\c-\C$, on the NIA one has
\beq
  \D \bar{G}^{\rm L} (-i\c) \approx
  \frac{2i a_2 \xi}{(\xi-b)^2 + c^2}\;,
\eeql{eq:fit1}
with $a_1 c + a_2 b = 0$ implementing the zero at $\c = \C$.  The broken line in Fig.~\ref{fig5}b shows this fit for $\ell=2$, yielding
\beq
 \w_\pm + i\C = \mp c - ib \approx  \mp 0.027 + 0.0033i\; .
\eeql{eq:position}
Further, a plot of $|J_+\bm{(}\zeta - i(\C {+} b)\bm{)}|$ vs.\ $\zeta>0$ (not shown) clearly supports a simple zero at $\zeta\approx\nobreak-c$ [cf.\ (\ref{eq:cutint04a})].

In summary, we have extrapolated Leaver's series to 
the unphysical sheet, revealing nearby
poles making the largest contribution
to the cut.  To the best of our knowledge, this
is the first time such QNMs have been found, and these
obviously have more effect on the dynamics than
QNMs on the physical sheet at larger $\left|\im\w\right|$.

In a broader context, consider the Kerr black hole.
By comparing numerics for moderately small rotation $a$~\cite{ono}
with the QNM multiplet found analytically to branch off
from $\C$ at infinitesimal~$a$, one concludes that
one \emph{additional} multiplet has to
emerge (as $a$ increases) near $\w = -i\C$.
Rather than the possibilities contemplated
in Ref.~\cite{Alec}, we speculate that this multiplet
may be due to the unconventional poles
discussed here splitting (as they must when spherical
symmetry is broken) and moving through the NIA as $a$ is 
tuned~\cite{rot}.

Thus it is advantageous to consider another Fourier contour
going into the unphysical sheet and detouring around~$\w_\pm$ (Fig.~\ref{fig1}b), including them as QNM contributions (line~b in Fig.~\ref{fig6}).  This slightly reduces the continuum (due to the integral along the NIA and often neglected as ``background"). More importantly, suppose these poles move through the NIA  as a parameter (say,~$a$) is tuned, becoming conventional QNMs.  In the assignment of Fig.~\ref{fig1}b, the total QNM and continuum contributions are \emph{separately} continuous; for the conventional contour, each is discontinuous.

%----------------------------------------------------------
\subsection{Poles on unphysical sheet: analytics}
\label{om*sec}

Interestingly, the above extrapolation can also be carried out analytically, by assuming the linearization of $J_+(\w{\approx}{-}i\C)$ to be valid up to the nearest zero,
\beq
  \w_++i\C\approx-\frac{J(-i\C)}{J'_+(-i\C)}\;.
\eeql{om-J}
Since $\D J(-i\C)$ vanishes, one does not have to indicate the sheet in the numerator. Following the methods and notation of~\cite{Alec}, one readily obtains
\beq\begin{split}
  J(-i\C)&=\frac{2\nu N}{2\nu+3}\left(\frac{\c_3}{\c_5}-1\right)\\
  &=-\frac{700009}{917504}\qquad\mbox{for $\ell=2$}\;,
\end{split}\eeql{JOm}
with the constants
\begin{gather}
\begin{aligned}[t]
  \gamma_3&\equiv
  -\Biggl[\hn\frac{9}{2}\sum_{j=0}^{N-2}\frac{N^{j+1}}{j!}
         +3(2\nu{+}3)\frac{N^N}{(N{-}1)!}\Biggr]e^{-N}\\
  &=-\frac{28\cdot11093}{3}e^{-8}\qquad\mbox{for $\ell=2$}\;,
\end{aligned}\label{def-g3}\displaybreak[0]\\
\begin{aligned}[t]
  \gamma_5&\equiv\frac{3\nu N^{N+1}e^{-N}}{(1-2\nu)(\nu+1)(N-2)!}\\
  &=-\frac{2^{24}}{135}e^{-8}\qquad\mbox{for $\ell=2$}
\end{aligned}\label{def-g5}
\end{gather}
characterizing $f(-i\C)$ and $g(-i\C)$ respectively (cf.\ Appendix~\ref{Jpr}). The peculiar integer in the numerator of (\ref{JOm}) has already surfaced in Ref.~\cite{Alec}.

Calculating $J'_+(-i\C)$ is formidable (Appendix~\ref{Jpr}), but since $\w_++i\C$ in (\ref{eq:position}) is mostly real, one can first evaluate only $\re J'_+$---much simpler due to the zero in $\D g$:
\begin{align}
  \re J'_+(-i\C)&=\half\D J'(-i\C)
  =-\half q'(\C)J\{g(i\C),f(-i\C)\}\notag\\
  &=-\frac{\c_3 q'(\C)}{4\nu(2\nu+3)}\;.\label{reJpr}
\end{align}
Ignoring the imaginary part, substitution yields
\begin{align}
  \w_++i\C&\approx\frac{8\nu^2\hn N}{q'(\C)}\left(\frac{1}{\c_5}-\frac{1}{\c_3}\right)\label{om*}\\
  &=-0.0328\ldots\qquad\mbox{for $\ell=2$}\;,\notag
\end{align}
comparing favorably with the numerical $-0.027$, where the latter value, also found by extrapolation from the physical sheet, need not be more accurate. Taking (\ref{reJpr}) as an approximation for $J'_+$ provides a bound: given that (\ref{om*}) has yielded an $\w_+$ close to $-i\C$, including $\im J'_+$ will render $\w_+$ even closer. Indeed, inserting into (\ref{om-J}) the \emph{exact} (\ref{Jres}) for $J'_+(-i\C,\ell{=}2)$ yields
\beq
  \w_++i\C\approx-0.03248+0.003436i
\eeql{om*full}
as our final estimate; especially the agreement of $\im\w_+$ with (\ref{eq:position}) is remarkable.

Combining (\ref{eq:cutint03}) and (\ref{eq:fit1}), our linearization of $J$ immediately yields $a_2\approx\re\boldsymbol{(}J'_+(-i\C)\boldsymbol{)}^{-1}\approx-\re\w_+/J(-i\C)\approx-0.0426$. The agreement with the numerical fit $a_2\approx-0.0227$ is slightly worse than the one for $\w_+$ itself.

For $\ell=3$, the same procedure would lead to $\w_++\nobreak i\C\approx-0.847+0.0433i$---outside the radius of convergence of the series for $J_+(\w)$, which is $\half$ due to its poles at the anomalous points of~$f$. Heuristically removing these poles by rescaling $J$, e.g., by linearizing $H(\w)\equiv\linebreak[1]J(\w)\*\sin(2\pi i\w)/(\w{+}i\C)$ [cf.\ (\ref{eq:norm}) below], does not change the prediction for~$\w_\pm$. Besides, the fact remains that $J(\w)$ has structure on a scale of $O(1)$. Thus, already for $\ell=3$, currently there is only slight evidence for $\w_\pm$ near $-i\C$. Similarly, for large $\ell$, the asymptotics of $q'(\C)$ given at the end of Section~\ref{cut-num} yield (\ref{om*}) as $\w_++i\C\sim-0.051\C$, far outside the permitted range of extrapolation. While we have not fully studied the large\nobreakdash-$\ell$ asymptotics of $\im J'_+(-i\C)$, this is unlikely to change the conclusion that our calculation is valid for $\ell=2$ and possibly $\ell=3$ only.

%----------------------------------------------------------
\subsection{QNMs versus background}
\label{subsect:compare}

In terms of $t'$, also the QNM contributions can be expressed for $x, -y \rightarrow \infty$ in a position-independent manner. The various magnitudes and phases are then completely determined---unlike the usual situation in (Q)NM analysis. Fig.~\ref{fig6}, line~c shows the contribution of the first QNM (i.e., the pair with the smallest damping, at $\omega_1 = \pm 0.747 - 0.178 i$), reduced by $10^3$. Insofar as the low QNMs are likely to be the most important for signal analysis, the comparison in Fig.~\ref{fig6} provides the estimate that the ``background" is small.

The above numerics refer to the most important case $\ell = 2$. The regions near the much larger $\C(\ell{\ge}3)$ are currently unattainable numerically.
In any event, Section~\ref{om*sec} shows that the unconventional poles, if present at all, must be further removed from $\C$, so their influence on the cut is likely to
be less pronounced.

%----------------------------------------------------------

\subsection{Zerilli equation}
\label{subsect:ze}

The corresponding quantity for the ZE is trivially $\D \tilde{\bar{G}}{}^{\rm L} (-i\c) = 
\D \bar{G}^{\rm L} (-i\c)$, by (\ref{eq:cutint04a}) and $\tilde{J}= J$~\cite{Wong1}.
Thus, the discussion of Fig.~\ref{fig6} also applies to the ZE.

Since the ZE has a QNM
at $-i\C$ but the RWE does not~\cite{Alec}, it might appear that the two $\bar{G}^{\rm L}$'s
should have different pole structures there. Indeed, $\tilde{\bar{G}}$ does have a pole, with residue $\propto\nobreak\tilde{f}(y,-i\C)\tilde{g}(x,-i\C)$.  However, $-i\C$ is anomalous for the ZE (but miraculous for the RWE), so $\tilde{f}$ does
not contain the normal growing solution, eliminating the residue in~$\tilde{\bar{G}}^{\rm L}$.

Physically, for a source \emph{not}
at the horizon, the axial (RWE) dynamics does \emph{not} have a term $e^{-\C t}$,
but the polar (ZE) dynamics does---with, however, vanishing amplitude as the source
approaches the horizon ($y \rightarrow -\infty$). 

%==========================================================

\section{Discussion}
\label{sect:discussion}

Like the hydrogen atom in quantum mechanics,
the Schwarzschild black hole is the simplest compact 
object in relativity, and its spectrum also  
contains a continuum in addition to the discrete part.
We have characterized this continuum, 
recovering the behavior both for
$\gamma \equiv i\omega \rightarrow 0$  
and near the miraculous point~$\C$. The present numerical method cannot access much larger~$\c$, 
but the pattern already appears clear. 

Although $\D \bar{G}^{\rm L}(-i\C)=0$, it surprisingly is largest \emph{near}~$\C$.
Moreover, for a limiting source and observer, the cut contribution
is small relative to the QNMs---of relevance once gravitational waves are detected.

Signal analysis in terms of a $V(x)$ is formally an inverse problem.
These are well-studied for closed systems~\cite{gelfand}:
\emph{two real} spectra determine 
$V$ on a finite interval. For open systems, we conjecture that 
\emph{one complex} spectrum suffices---provided the discrete
QNMs are complete. The cut (rendering them incomplete) is likely
to hamper inversion, but, intriguingly,
the \emph{extended} family of QNMs (including the unconventional ones) \emph{may} be complete
for the dynamics (apart from 
the prompt signal) and permit inversion, 
even though $V$ is not
finitely supported~\cite{Ching-QNM}.  
The present at least shows that \emph{one} pair of nearby poles on the unphysical sheet already dominates the cut.

These questions may be explored through solvable
models with potential tails~\cite{Ferrari,Cardoso}. Some aspects of the RWE and ZE can also be analyzed asymptotically~\cite{Alec2}.  
Numerical algorithms
(e.g., generalizing the continued-fraction
method~\cite{Leaver}) valid on the NIA and even into the unphysical region would also be
useful, allowing QNMs there to be studied directly rather than through extrapolation.

Finally, it would be instructive to evaluate
the continuum for a Kerr hole, to
study any unconventional poles, in particular 
their movement and possible emergence onto the
physical sheet as the rotation is increased from zero~\cite{berti}.  

%==========================================================

\acknowledgments
We thank E. Berti, E.S.C. Ching, Y.T. Liu, W.M. Suen and 
C.W. Wong for many discussions, and the Hong Kong Research Grants
Council for support (CUHK 4006/98P). AMB was also supported by a C.N. Yang Fellowship.

%==========================================================

\appendix

%==========================================================

\section{Leaver's series}
\label{app:NA}

Suppressing the parameter $\w$, we first calculate an \emph{un}normalized $\underline{g}$. Define $R\equiv g'\!/g=\underline{g}'\!/\underline{g}$. Choose $z$ and compute $R(z)$ by Miller's algorithm of downward recursion on Leaver's series~\cite{Leaver series,YTLiu}. With, say, $\underline{g}(z)\equiv1$, $\underline{g}(x)$ is then integrated trivially for all~$x$. Thus, the norm $\mathcal{N}=\lim_{x\To \infty}e^{i\w x}\!/\underline{g}(x)$ is not CH; convergence provides a check. Finally, $g=\mathcal{N}\underline{g}$. This procedure requires only \emph{one} CH calculation [for $R(z)$], instead of one for each $x$.

In principle, the result should be $z$-independent; in practice we take $|z| \sim 1$. However, we have checked that the $R(z)$ thus obtained obey $R'=V-\w^2-R^2$. Such tests show that Miller's algorithm works well at least for $-4.25< \im \w <-1.25$, 
$10^{-4}< \re \w <0.1$, and $-0.5<z<1.5$; by optimizing $z$, the range on the NIA can be slightly extended to $0.5 \lesssim \gamma \lesssim 6.0$,
varying somewhat with $\ell$.
One also verifies that $R(z{\To}{\pm}\infty)\To\nobreak\pm i\w$.

In practice we calculate $q(\c)$ using a small $\e \neq 0$ and some definite $x$ in $(-5,0)$; the smaller $\e$, the wider the range of permissible~$x$. Convergence for $\e \To 0$ is rapid, and the $x$-independence in particular gives an accuracy estimate. On this basis, the decimals given in the main text should be significant, and the error bars in the figures should be tiny on the given scale.

%==========================================================
\section{Jaff\'e's series}
\label{app:jaffe}

This appendix presents Jaff\'e's series~\cite{Leaver series,Jaffe},
with particular attention to the anomalous and miraculous
points. The series, first used for
the $\mathrm{H}_2^+$ ion and applied to the
present problem by Leaver~\cite{Leaver series}, expresses the outgoing solution, $f(r{\downarrow}1,\w) \sim e^{-i\w x} = (r{-}1)^{-i\w}e^{-i\w r}$, as
\beq
  f(r,\w)=(r{-}1)^{-i\w}r^{2i\w}e^{i\w r} \sum_{n=0}^{\infty}a_{n} (\w)
  \left(\frac{r{-}1}{r} \right)^{n} \;.\label{eq: Jaffe}
\eeq
The coefficients $a_{n}(\w)$ satisfy
\beq
  \a_{n}a_{n+1}+\b_{n}a_{n}+\c_{n}a_{n-1}=0 \; ,
  \label{eq: recurrence of a_n}
\eeq
$n = 1, 2, \ldots$, with $a_{n<0}=0$ and ($s=2$ throughout)
\begin{align}
  \a_n&=(n{+}1)(n{+}1{-}2i\w) \;, \notag\\
  \b_n&=-2n^{2}+(8i\w{-}2)n+8\w^{2}+4i\w-\ell(\ell{+}1)+3 \;, \notag\\
  \c_n&=n^{2}-4i\w n-4\w^{2}-4 \;.
\end{align}
The normalization $a_{0}$ can be  chosen freely (see below).

Jaff\'e's series leads to some important insights. 
First, $\a_{n}$ has a zero at $\w=-i(n{+}1)/2$.
So at $\w_n =-in/2$ ($n=1,2,\ldots$),
$a_{m\ge n}$ and  
therefore $f(\w)$ generically have simple poles,
removable through scaling by $\w - \w_n$, in effect killing
the $a_{m<n}$ and making $a_{m\ge n}$  
finite~\cite{Alec}.  Then the leading behavior is $\chi(r, \w_n) \sim (r{-}1)^{n/2} \sim e^{i \w_n x}$, an \emph{incoming} wave. Thus, at these anomalous 
points, $f\propto\chi$ is outgoing
(being the analytic continuation of outgoing waves defined
for $\im \w >0$) \emph{and} incoming.

However, according to (\ref{eq: recurrence of a_n}), $a_{n}$ 
(hence, all $a_{m\ge n}$) can 
be non-singular even when $\a_{n{-}1}=0$, 
if $\b_{n{-}1}a_{n{-}1} + \c_{n{-}1}a_{n{-}2} = 0$. 
This is called a miracle; in the present case
for $\ell =2$, it occurs at $n = 8$, i.e., at $\w = -i\Gamma$~\cite{Alec}.

Because of the poles at the anomalous but not at
the miraculous $\w_n$, we choose to calculate $\chi$ with
\beq
  a_0 = \frac{ \sin 2\pi \gamma } { \gamma - \Gamma } \; ,
\eeql{eq:norm}
in which case all functions are nonsingular at any $\w_n$ (both anomalous and
miraculous). However, if needed [e.g.\ in (\ref{eq:cutint04a})] we can always find the normalization by taking $r\downarrow1$; cf.\ Appendix~\ref{app:NA}.

Finally, note that $f(\w)$ has no cut in the $\w$\nobreakdash-plane,
as is also evident from more general considerations.

%==========================================================

\section{Frequency derivative of the Wronskian}
\label{Jpr}

\subsection{Preliminaries}

Writing $f_1\equiv\partial_\w {f(\w)|}_{-i\C}$ and similar for $g$, we are interested in (arguments $-i\C$ suppressed where possible)
\beq
  J'_+=g_{1+}d_xf+gd_xf_1-f_1d_xg-fd_xg_{1+}\;.
\eeql{defJpr}
The differentiated RWE
\beq
  [d_x^2-\C^2-V(x)]f_1(x)=2i\C f(x)
\eeql{diff-WE}
etc., plus nontrivial differentiated OWCs, determines $f_1$ and $g_{1+}$. The calculation proceeds in $r$, using $x(r)$ as a shorthand only. One introduces the SUSY generator
\beq
  \xi_1(r)=\frac{2\nu r+3}{r}\hp e^{-Nx/2}
\eeql{def-xi1}
and the secondary solution $\xi_2(r)=\xi_1(r)\int_1^r\!dt\,[t/(t{-}1)]\*\xi_1^{-2}(t)$, where the integral is elementary~\cite{Alec}:
\beq\begin{split}
  \frac{\xi_2(r)}{\xi_1(r)}
  &=\frac{1}{\c_5}\biggl[e^{N(r-1)}\sum_{j=0}^{N-2}
    \frac{[N(1{-}r)]^j}{j!}-1\biggr]\\
  &\quad+\frac{e^{Nr}(r{-}1)^{N-1}[2\nu r^2-(2\nu{+}3)r+6]}{4\nu^2N(2\nu r+3)}\;.
\end{split}\eeql{xi2}
Note that the decay of $\xi_2$ near the horizon is due to a high-order cancellation, cf.\ the first line of~(\ref{xi2}). This behavior complicates the entire calculation, right down to the floating-point evaluation of (\ref{Jres}) or its higher-$\ell$ counterparts. Namely, for $\ell=2$ ($\ell=3$) the first 3 (16) digits cancel in the rational and transcendental [largely $\propto\nobreak\Ei(N)$] contributions to $\im J_+'$, so that very high accuracy is needed already for moderate $\ell$.

The Wronskian is readily evaluated as
\beq
  J(\xi_1,\xi_2)=1\;.
\eeql{J12}
With $\xi_j\equiv\xi_1+\c_j\xi_2$, the outgoing functions are $f=\xi_3/(2\nu+3)$ and $g=2\nu N\xi_5/\c_5$, while the incoming ones read $f(i\C)=(2\nu+3)N\xi_2$ and $g(i\C)=\xi_1/(2\nu)$; the latter has been used already in deriving the final line of (\ref{reJpr}).

\subsection{Differentiated wave functions}

Using a Green's function approach, (\ref{diff-WE}) is solved as
\beq\begin{split}
  f_1(r)=\frac{iN}{2\nu+3}\biggl[&\xi_2(r)\!\int^r\!dt\,\frac{t\xi_1(t)\xi_3(t)}{t-1}\\
    &-\xi_1(r)\!\int^r\!dt\,\frac{t\xi_2(t)\xi_3(t)}{t-1}\biggr]\;,
\end{split}\eeql{f1a}
where the undetermined integration constants reflect the possibility of adding a homogeneous solution to (\ref{diff-WE}). Of course, $f_1$ itself is not arbitrary, and the second of these constants follows from $f_1(x{\To}{-}\infty)\sim[-ix+O(e^x)]e^{-\C x}$. The required asymptotics of the integrals in (\ref{f1a}) are straightforward; for definiteness we write the answer as
\begin{gather}
\begin{aligned}[b]
 iN^{-1}(2\nu{+}3)f_1&=
    \xi_1\{B+N^{-1}x+\c_3D-N^{-2}\}\\
    &\quad-\xi_2\{C+\c_3[B{+}N^{-1}x]+\beta\};
\end{aligned}\label{f1b}\displaybreak[0]\\
\begin{split}
  B(r)&=\int_1^r\!dt\,\frac{t}{t-1}\hp[\xi_1(t)\xi_2(t)-N^{-1}]\;,\\
  C(r)&=\int_\infty^r\!dt\,\frac{t\xi_1^2(t)}{t-1}\;,\qquad
  D(r)=\int_1^r\!dt\,\frac{t\xi_2^2(t)}{t-1}\;.
\end{split}\label{aux}
\end{gather}
The integrals~(\ref{aux}) can be evaluated by partial-fraction expansion, yielding only elementary functions plus $\Ei(r)\equiv\int_{-\infty}^rdt\,e^t\!/t$~\cite{lookout}. However, the full primitives are cumbersome---not surprising since already $\xi_2$ is involved. Hence, we only give results when required.

One cannnot find $\beta$ from low-order asymptotics, as was the case for $\c_3$ in the calculation of $f$ itself~\cite{Alec}. For $\ell=2$, we have obtained the Born series for $f(r{\approx}1,\w)$ by computer algebra. Only after $\partial_\w$ is taken in each term does one set $\w=-i\C$, comparing the result to the analogous expansion of (\ref{f1b}). The first eight terms agree, as they must. The ninth [$O\boldsymbol{(}(r{-}1)^4\boldsymbol{)}$] terms agree if
\beq
  \beta=9\Ei(-8)+\frac{\c_3}{4}(\c_{\rm E}{+}\ln8)+\frac{4226209}{60}\hp e^{-8}\;,
\eeql{beta2}
and for this value the two series coincide. The transcendental $\Ei$, $\c_{\rm E}$, and $\ln8$ only occur to cancel their counterparts in the expansion of $C(r)$, defined with a lower limit $r=\infty$ to ensure convergence---the Born series for $f_1$ essentially involves rationals only. Indeed, $\Ei(-8)$ does not occur in (\ref{Jres}) below.

For general $\ell$, we calculate $\beta$ by requiring that $\partial_\w^2{[(r-1)^{i\w}f(r,\w)]}_{-i\C}$ be single-valued near $r=1$. This generalizes the determination of $\c_3$ and hence $f$ in~\cite{Alec}, by demanding that $f_1(r)+if(r)\ln(r-1)$ be a power series near $r=1$. Using the latter, presently we find that
\beq\begin{split}
 &f_2(r)-2iN^{-1}f_1(r)+2if_1(r)\ln(r{-}1)\\
 &+[2N^{-1}-\ln(r{-}1)]f(r)\ln(r{-}1)
\end{split}\eeql{no-log}
should not involve $\ln(r{-}1)$. The equation for $f_2$ analogous to (\ref{diff-WE}) yields [cf.\ (\ref{f1a})]
\begin{multline}
  f_2(r)-\frac{2i}{N}f_1(r)=2iN\biggl[\xi_2(r)\!\int^r\!dt\,\frac{t\xi_1(t)f_1(t)}{t-1}\\
    -\xi_1(r)\!\int^r\!dt\,\frac{t\xi_2(t)f_1(t)}{t-1}\biggr]\;.
\end{multline}
Insertion into (\ref{no-log}) using (\ref{f1b}) for $f_1$ throughout eventually yields that all logarithms indeed cancel if
\begin{align}
  \beta&=9\Ei(-N)-\frac{(18\nu{+}9)\c_5}{4\nu^2N^2}
  +\frac{2\c_3}{N}\!\left[\c_{\rm E}{+}\ln N{+}\frac{1}{N}\right]\notag\\
  &\quad+\sum_{j=1}^{N-1}\frac{9}{j}
  \biggl[e^{-N}\sum_{k=j}^{N-2}\frac{N^k}{k!}-\frac{\c_5}{2\nu^2N}\biggr]+N\hn\rho\;.
\label{beta-res}\end{align}
The residue $\rho\equiv\res[r\xi_1^2(r)B(r)/(r{-}1)]_{r=1}$ reads
\beq
  \rho=(2\nu{+}3)^2c_N+(9{-}4\nu^2)c_{N-1}+9\sum_{p=1}^{N-2}c_p\;,
\eeql{rho}
\begin{widetext}
\beq\begin{split}
  \frac{\c_5e^{2N}c_p}{N^p}&=2\c_3e^N
  \sum_{j=1}^p\biggl(\frac{N^j}{j!}-1\biggr)\frac{N^{-j-1}}{j(p{-}j)!}
  -\sum_{j=1}^p\sum_{k=0}^{j-1}\frac{9N^{k-j}}{j(p{-}j)!k!}
  +\frac{(2\nu{+}3)^2N^{N-1}}{(N{+}p)!}(2^{N+p}{-}1)\\
  &\quad-\frac{(2\nu{+}3)^2N^{N-2}}{p!(N{-}1)!}
    +\sum_{j=1}^{N-1}\left(\frac{(j{-}1)!2^{p+j}}{(p{+}j)!}-\frac{1}{jp!}\right)
    \biggl\{9\sum_{k=j}^{N-2}\frac{N^k}{k!}+6(2\nu{+}3)\frac{N^{N-1}}{(N{-}1)!}\biggr\}\\
  &\quad+\sum_{j=1}^{N-1}\frac{1}{j(p{+}j)!}\biggl\{\frac{4\nu^2N^{N-1}}{(N{-}j{-}1)!}
  -9\sum_{k=j}^{N-1}\frac{N^k}{(k{-}j)!}-\frac{(2\nu{+}3)^2N^N}{(N{-}j)!}\biggr\}\;.
\end{split}\eeql{cp}
For $\ell=2$, (\ref{beta-res})--(\ref{cp}) are verified to reduce to (\ref{beta2}).
\end{widetext}

The calculation of $g_{1+}$ may seem more difficult conceptually, involving analytic continuation in~$r$. It actually is easier technically, since the exact form \emph{does} follow from low-order asymptotics. The counterpart to (\ref{f1a}) reads
\beq
  \frac{\c_5^2g_1(r)}{2i\nu N^2}=\xi_5(r)\!\int^r\!dt\,\frac{t\xi_1(t)\xi_5(t)}{t-1}
    -\xi_1(r)\!\int_{-\infty}^r\!dt\,\frac{t\xi_5^2(t)}{t-1}\;,
\eeql{g1a}
where the lower limit for the second term ensures $g_1(r{\To}{-}\infty)\To0$. We select $g_{1+}$ by continuing $r$ from $-\infty$ to the physical $r>1$ through the upper half plane. The remaining integration constant is made explicit as
\begin{align}
  \frac{\c_5^2g_1(r)}{2i\nu N^2}&=
  \xi_5(r)\biggl[\int_{-\infty}^r\!\frac{dt\,t}{t{-}1}\left(
    \xi_1(t)\xi_5(t)-\frac{\c_5}{N}\right)+\frac{\c_5x}{N}+\alpha\biggr]\notag\\
    &\quad-\xi_1(r)\!\int_{-\infty}^r\!dt\,\frac{t\xi_5^2(t)}{t{-}1}\;.
\end{align}
Comparison to the differentiated asymptotics of $g(\w)$,
\beq
  \frac{g_1(r)}{ie^{Nx/2}}=x-\frac{3x}{2\nu r}+\frac{3x}{Nr^2}+\frac{3}{\nu Nr}+O(r^{-2})\;,
\eeql{g1b}
yields $\alpha=\c_5/(4n^2N^2)$ in $O(r^0)$ of the last factor in (\ref{g1b}); agreement of $O(r^{-1})$ provides a check.

\subsection{Evaluation of the differentiated Wronskian}

Substituting (\ref{f1b}) and (\ref{g1b}) into (\ref{defJpr}), the prefactors of the $t$-integrals are all handled using (\ref{J12}), yielding
\begin{align}
  \frac{(2\nu{+}3)\c_5}{2i\nu N^2}J'&=
    \int_{-\infty}^r\!\frac{dt\,t}{t{-}1}\!\left[\frac{\c_5{-}\c_3}{N}-\xi_3(t)\xi_5(t)\right]
    +\frac{\c_3{-}\c_5}{4\nu^2N^2}\notag\\ &\quad-\c_5N^{-2}+\beta
    +C(r)+(\c_3{+}\c_5)B(r)\notag\\ &\quad+\c_3\c_5D(r)+2\c_3N^{-1}x\;.\label{Jint}
\end{align}
Differentiating with respect to $r$, the $r$-independence of (\ref{Jint}) is readily verified.  Indeed, the rhs is merely a regularized $\int_1^{-\infty}dt\,\xi_3\xi_5t/(t{-}1)$, which unfortunately diverges at both ends. Choosing $r\To-\infty$, the first term vanishes.

Inspecting (\ref{aux}), it is gratifying that the individual exponential divergences (for $r\To-\infty$) in $B$, $C$, and $D$ indeed cancel in (\ref{Jint}). Taking the $t$-contours in the upper half plane fixes the contributions of the singularities at $t=0,1$, yielding $\re J'_+$ as in (\ref{reJpr}). In full, one has
\begin{widetext}
\begin{multline}
  \lim_{r\To-\infty+i\eta}\left[C(r)+(\c_3{+}\c_5)B(r)+\c_3\c_5D(r)+2\c_3N^{-1}x\right]=\\
  \begin{aligned}[b]
  &\frac{2\c_3}{N}\!\left(\frac{\c_3}{\c_5}-2\right)\!(\c_{\rm E}{+}\ln N)
  -9\Ei(-N)+\frac{4\c_3^3}{9N^2\c_5}\Ei(N{+}i\eta)
  +2\pi i\frac{\c_3}{N}\!\left(1-\frac{\c_3}{\c_5}\right)\\
  &+\sum_{j=1}^{N-1}\frac{9}{j}
  \biggl[\frac{\c_5}{2\nu^2N}-e^{-N}\sum_{k=j}^{N-2}\frac{N^k}{k!}\biggr]
  +\frac{3\c_3(2^N{-}1)}{(2\nu{+}3)N^2}
  +\frac{2(2\nu{+}3)(2\nu{+}5)(8\nu^2{+}8\nu{-}3)\c_3}{9N^2(N{-}1)(N{-}2)}\\
  &+\frac{2\c_3}{N^2}+\frac{(9{-}6\nu)\c_3}{2\nu^2}\frac{N{-}3}{N{-}2}
  -\frac{9\c_3\c_5e^N}{4\nu^4}\sum_{j=0}^{N-4}\frac{j!}{N^{j+3}}
  -\frac{(12\nu{+}9)\c_3N^Ne^{-N}}{\c_5}\sum_{j=1}^N\frac{1}{jj!(N{-}j)!}\\
  &+\frac{3(2\nu{+}3)\c_5}{4\nu^2N^2}+\frac{2\c_3}{N}\sum_{j=1}^{N-2}\frac{1}{j}
  -\frac{9\c_3e^{-N}}{\c_5}\sum_{j=1}^{N-1}\sum_{k=1}^j\frac{N^j}{kk!(j{-}k)!}
  +\frac{9\c_3}{\nu^2}\sum_{j=3}^{N-2}\sum_{k=0}^{j-3}\frac{N^{j-k-2}k!}{j!}\\
  &+\frac{4\nu\c_3e^{-N}}{\c_5}\sum_{j=0}^{N-4}j![(\nu{+}3)N^N\sigma_{j+3}
  -\nu N^{N-1}\sigma_{j+2}]
  -\frac{9\c_3e^{-N}}{\c_5}\sum_{j=1}^{N-5}N^{N+j}\sigma_{j+3}\sum_{k=0}^{j-1}\frac{k!}{N^k}\;,
  \end{aligned}\\[-1cm]\label{lim-res}
\end{multline}
\end{widetext}
with $\sigma_j\equiv\sum_{k=j}^{N-2}{[k!(N{-}2{+}j{-}k)!]}^{-1}$. The infinitesimal $\eta>0$ selects the upper branch of $\Ei(N{+}i\eta)$.

Finally, Eqs.\ (\ref{def-g3}), (\ref{def-g5}), (\ref{beta-res})--(\ref{cp}), and (\ref{Jint})--(\ref{lim-res}) combined yield a tedious, but explicit exact result for $J'_+$; some cancellation occurs. For $\ell=2$, one has
\pagebreak
\begin{align}
  J'_+=\frac{i}{49\cdot2^{38}}\bigl[&{-}17122265640585(\c_{\rm E}{+}\ln8{-}i\pi)\notag\\
    &-245810518235861775\Ei(8{+}i\eta)e^{-8}\notag\\
    &+36326230655979688\bigr]\;.\label{Jres}
\end{align}

%==========================================================


\begin{thebibliography}{99}
\bibitem{q'}
One has $q'(\C)=-8\pi\nu^2\*[2N(\c_3{-}\c_5)/\c_5^2+4\c_3^2/9\c_5^2]$, with $\c_3$, $\c_5$ as in (\ref{def-g3}) and (\ref{def-g5}) respectively; see (8.2) in~\cite{Alec}. Since $\c_5\sim-\sqrt{3/2\pi}N^2$ while $\c_3=O(N)$ for $\ell\To\infty$, the latter can be neglected asymptotically, yielding the result at the end of Section~\ref{cut-num}.
\end{thebibliography}
\end{document}